# Microwave AC voltage induced phase change in Sb₂Te₃ nanowires

*Pok-Lam Tse, Laura Mugica-Sanchez, Fugu Tian, Oliver Rüger, Andreas Undisz, George Moethrath, Susumu Takahashi, Carsten Ronning and Jia Grace Lu\**

P. L. Tse
Department of Chemical Engineering and Materials Science, University of Southern California, Los Angeles, California, CA 90089, USA

L. Mugica-Sanchez and Prof. Dr. S. Takahashi
Department of Chemistry, University of Southern California, Los Angeles, California, CA 90089, USA

O. Rüger and Prof. Dr. C. Ronning
Institute of Solid State Physics, Friedrich Schiller University of Jena, 07443 Jena, Germany

Dr. A. Undisz
Otto Schott Institute of Materials Research, Friedrich Schiller University of Jena, 07443 Jena, Germany

F. Tian
Department of Electrical and Computer Engineering - Electrophysics, University of Southern California, Los Angeles, CA 90089, USA

G. Möthrath and Prof. Dr. J. G. Lu
Department of Physics and Astronomy, University of Southern California, Los Angeles, California, CA 90089, USA
Department of Electrical and Computer Engineering - Electrophysics, University of Southern California, Los Angeles, CA 90089, USA

*Corresponding author, E-Mail: jialu@usc.edu

## Abstract

Scaling information bits to ever smaller dimensions is a dominant drive for information technology (IT). Nanostructured phase change material emerges as a key player in the current green-IT endeavor with low power consumption, functional modularity and promising scalability. In this work, we present the demonstration of microwave AC voltage induced phase change phenomenon at ~3 GHz in single Sb₂Te₃ nanowires. The resistance change by a total of 6 – 7 orders of magnitude is evidenced by a transition from the crystalline metallic to the amorphous semiconducting phase, which is cross-examined by temperature dependent transport measurement and high-resolution electron microscopy analysis. This discovery could potentially tailor multi-state information bit encoding and discrimination along a single nanowire, rendering technology advancement for neuro-inspired computing devices.

## 1. Introduction

Non-volatile memory cells with high read/write speed have been the quest in the past decades for superior performance in computing and communication electronics.[1-3] Phase change materials (PCM) have evolved into the industrial development stage when the traditional technologies decelerate. PCM demonstrates a competitive read/write speed compared to flash and embedded memories in solid state drives.[4-6] With their high speed and remarkable cycling endurance, PCM in nanostructured forms is considered as a potential replacement for storage-class memory with exceptional computing performance and reduced power consumption.

$Ge_2Sb_2Te_5$ has been extensively studied by slow passive heating and annealing[2,7] to judiciously separate out the amorphous, cubic and hexagonal crystalline phases.[8,9] In comparison, the binary $Sb_2Te_3$ compound has similar phase change properties but a lower melting temperature[10] than that of $Ge_2Sb_2Te_5$. This implies reduced power requirement for information encoding, and thus relieves the heating problem on an integrated device chip. In addition, $Sb_2Te_3$ attracts increasing research interests due to the existence of linearly dispersed surface states[11] in the band structure, which could bring new perspectives from quantum mechanics in future device designs.

Conventional operation of phase change random access memory (PCRAM) is based on the transition between the crystalline and amorphous phase, which can yield several orders of magnitude difference in the electrical resistance.[4,8,12] In such a design, applying a direct current (DC) voltage pulse in a single step to memory cells is the most common method to switch memory states.[8,13] In our work, we present alternate current (AC) voltage sweep measurements from radio frequency to microwave range on $Sb_2Te_3$ nanowires, which reveal a systematic stepwise increase in DC resistance at ~3 GHz. The samples are cross-investigated by transport measurements, as well as high resolution electron microscopy analysis on segments with and without undergoing the AC voltage sweeps.

This unique phenomenon suggests that one can change the resistance with an AC frequency knob. This extra degree of freedom for memory state switching can lead to innovative developments in PCRAM for neuromorphic computation. It provides a futuristic technique to control the amount of phase change in the material, which can be utilized to obtain the optimum crystalline-amorphous ratio in order to reach a targeted resistance as an intermediate memory state. Consequently, such intermediate states pave the way to a new frame of applications of multi-level storage with a simple device, which is very difficult to be achieved with the traditional heating or application of voltage pulses.

## 2. Experimental Methods

$Sb_2Te_3$ nanowires (NWs) were synthesized by Au-catalyzed chemical vaporized deposition (CVD) enabling the vapor-liquid-solid growth mechanism.[14] First, the $Si/SiO_2$ substrate was coated with 0.1 wt% poly-L-lysine to enhance the Au nanoparticle adhesion, then a solution with Au particles having diameters of about 30 nm was dispersed onto the substrate surface. Afterward, the CVD process was carried out in a quartz tube inside a horizontal furnace setup. The source material of 0.6 g Sb powder was placed in the center of furnace, 1 g Te powder at 13.5 cm away from center at upstream; and the Au nanoparticle coated $Si/SiO_2$ substrate was set at 10.5 cm downstream from center. Argon was used as a carrier gas at a flow rate of 80 standard cubic centimeters per minute (sccm) at 2.67 - 5.33 mbar partial pressure. The process began with heating up the quartz tube to 430 °C in 20 minutes, then maintaining the temperature for 6 hours. Pristine NWs on the as-grown substrates were then transferred to new substrates by sonication in isopropyl alcohol, and subsequently pipetted onto proper substrates for device fabrication and microscopy analysis. The NW devices with 100 nm Au electrodes on a 5 nm Ti adhesion layer were fabricated by photolithography and the device with Nb electrodes was fabricated by e-beam lithography.

In the AC sweep measurements, the source and drain contacts of the devices were connected to bias-Ts at the source electrode to mix DC signals from a DC analyzer (Agilent hp4156p) and AC signals from a microwave (MW) source (SRS SG380); and at the drain electrode, to separate DC signals to ground and AC signals to MW power detector (Windfreak SynthNV). **Figure 1** shows the circuit diagram of one NW device on a $Si/SiO_2$ substrate. Coaxial cables were used to minimize AC signal loss and the power detector was in place to ensure minimum MW power transmission at -49 dBm ($1.26 \times 10^{-8}$ W).

The device was cooled down to about 77 K in high-vacuum ($1.0 \times 10^{-7}$ mbar) in order to ensure that thermal effects were only dependent on the MW input power. The AC sweeps were conducted between 10 MHz and 4 GHz at 10 to 20 MHz steps; *I-V* sweeps from -150 nA to +150 nA were carried out at each frequency step, and the differential resistance was calculated for the nanowire at each corresponding frequency. Both forward and backward sweeps between 10 MHz and 4 GHz were conducted to probe the reversibility of the MW responses.

In order to investigate the temperature dependence of the electrical resistance, measurements were done during cooling down and warming up from 75 to 300 K. A reference pristine sample (NW4), without undergoing any AC sweep, was measured separately in the same temperature range to compare the temperature dependence of the resistance.

For electron microscopy analysis, cross-sections were taken from one single nanowire at two different parts: one segment was subject to AC sweeps, whereas the second segment did not. The specimen were prepared using a focused ion beam system. Briefly, a Pt protective layer was first deposited on the targeted NW segment by electron beam induced decomposition of molecules. Then the substrate material in the surrounding area was sputtered away using the focused ion beam. After that, the lamella was lifted off from the substrate and transferred to the omni-probe copper grid. Finally, the thickness of the lamella was thinned down to about 50 nm for high-resolution transmission electron microscopy (TEM) analysis using a CS-corrected JEOL NEOARM 200 F. The TEM analysis was performed with 80 keV electrons at low currents in order not to induce any phase change or degradation by the impact of electron beam, which was correspondingly checked upon the analysis.

## 3. Results and Discussions

$Sb_2Te_3$ has a bulk band gap of 0.28 eV and simple surface states consisting of a single Dirac cone in the band gap. The pristine crystalline structure of $Sb_2Te_3$ thin films and nanowires is hexagonal, and the primitive cell is rhombohedral ($R\overline{3}m$). Our previous studies on the nanowires have revealed the single crystalline structure with repeating quintuple layers of (Te-Sb-Te-Sb-Te) with an interlayer distance of 0.309 nm.[11,15-17] We have also in the past performed low temperature magnetoresistance measurements and angle resolved photoemission spectroscopy on the nanowires synthesized by the same setup as presented in this work. The observed periodic Aharonov-Bohm type oscillations are attributed to transport in topologically protected surface states in the $p$-type $Sb_2Te_3$ nanowires, with a Fermi level that situates around 40 meV below the $\Gamma$-point.[11]

The scanning electron microscope (SEM) image in **Figure 2a** shows a single $Sb_2Te_3$ nanowire on a $Si/SiO_2$ substrate. Energy dispersive X-ray (EDX) spectroscopy mapping for Sb and Te are displayed in **Figure 2b** and **Figure 2c**, respectively. The mappings of Sb and Te illustrate uniform elemental distributions along the nanowire. From the EDX spectra in **Figure 2d**, the atomic ratio of Sb:Te is calculated to be 2:3. Powder X-ray diffraction (XRD) (Rigaku Ultima IV diffractometer) is carried out in $\theta/2\theta$ mode with a scan speed of 4°/min. The crystal structure is confirmed to be rhombohedral (PDF # 00-015-0874), as shown in **Figure 2e**. A TEM image of the top view of the nanowire is displayed in **Figure 2f**, showing that the growth direction is along [110]. The lower right inset depicts the corresponding SAED pattern, verifying its single crystalline hexagonal/rhombohedral nature. The upper left inset of **Figure 2f** is the cross-section TEM image of a NW, manifesting the repeating quintuple layers (QL).

Three NW samples have been investigated at 77 K for MW responses, with their results plotted separately in **Figure 3**. Clearly, the common phenomenon for all three NWs is the increase in the electrical resistance sharply around 3 GHz, independent of the geometry or material of the electrodes contacting the NWs. NW1 with Nb electrodes at 1 µm apart has an initial resistance of about 2800 Ω. Its AC sweep measurement is plotted in **Figure 3a,** indicating a resistance jump to 3350 Ω at 3 GHz in the first forward sweep from 10 MHz to 4 GHz. Similar results are observed in NW2 (**Figure 3b**) and in NW3 (**Figure 3c**), both fabricated with Ti/Au source and drain electrodes at 2 µm apart. It has been shown from the comparison of 2-probe and 4-probe measurement that the contact between either Nb or Ti/Au to the nanowire is of ohmic nature with negligible contact resistance (data not shown).

NW2 sample was selected for subsequent continuous backward and forward AC sweeps at a power level around -30 dBm (1 µW) after the initial forward sweep displayed in **Figure 4**. For the first three sweeps shown (backward in black line, forward in red line and backward in blue line), the resistance starts to slightly increase between 2.8 and 3 GHz for both forward and backward direction. Then at the subsequent sweeps, the resistance change becomes much more pronounced when the frequency reaches 3 GHz. The final sweep spikes the resistance to a saturation level of $10^8$ Ω at 3 GHz, as illustrated in the green line.

NW2 was further investigated for the resistance change by passive heating to verify that the resistance change originated from a phase change and not due to other artifacts such as oxidation or decomposition. The results are shown in **Figure S1** in the supporting information. The reversibility of high and low resistances in alternating short and long heating steps verifies that the sample after AC sweeps had gone through phase transition.

After the two samples (NW2 & NW3) reached the saturation levels, the resistances were measured from 77 K to room temperature. It was found that the resistances of both samples first increased and peaked at ~100 K (as shown in **Figure 5a, 5b**), then they decreased exponentially with increasing temperature. Arrhenius semi-log fittings of *ln R* vs *1/($k_BT$)* (where *R* is DC electrical resistance, $k_B$ is Boltzmann constant, and $T$ is temperature), between 110 K and 300 K are plotted, as shown in the insets. The results reveal that the transport is governed by thermal excitation conduction with two distinct regions between 110 – 220 K and 220 – 300 K, and with respective activation energy around 48 meV and 100 meV. These values are close to the thermal activation energies attributed to phonon-assisted hopping of small polarons, *i.e.* localized charges that are "self-trapped" within potential wells produced by distorting the surrounding atoms.[18] The exact transport mechanism is by itself a fascinating topic, which requires additional studies with Seebeck and Hall measurements. In contrast to NW2 and NW3 samples, the pristine reference sample NW4, which did not undertake microwave sweeps, keeps a linear relation of resistance-temperature ($R$–$T$) with $R(\Omega) \propto 0.95(\Omega/K)\cdot T(K)$, as shown in **Figure 5c**, manifesting a metallic behavior as in the initial state with a typical temperature coefficient estimated to be ~0.0014 /K.

The next investigation was to find out whether one can achieve segment-wise encoding along one nanowire sample, *i.e.* to obtain different resistance states in desired segments of one single wire. Four electrodes were fabricated across a long nanowire over 10 µms (NW5, as shown in the inset of **Figure 6a**). The segment of NW between contacts 2-3 undertook AC sweeps to get switched from low resistance to high resistance state; whereas the segment between electrodes 4-5 was afloat, not connected to either the DC analyzer or the MW power source. **Figure 6a** shows the AC sweeping results of the segment between electrodes 2-3 at 77 K. The first sweep at -26 dBm MW power did not induce a resistance change at 3 GHz, whereas the second sweep at -24 dBm MW power increased the resistance sharply at 3 GHz. After continuous forward and backward sweeping at 77 K at -24 dBm, the NW reached and saturated at a high resistance state. Additional forward and backward AC sweeps did not yield any further resistance changes (see **Figure S2** in the supporting information). The $R$–$T$ relation of this segment between electrodes 2-3 is plotted in **Figure 6b** and **Figure 6c** upon cooling down and upon warming up after the AC sweeps, respectively. The linear relation during cool down indicates that the initial low resistance state has metallic behavior, consistent with the reference NW4 sample shown in **Figure 3c**. After reaching the high resistance state, the resistance decays exponentially with increasing temperature (**Figure 6c**). The inset in **Figure 6c** shows the Arrhenius semi-log fitting, confirming the two ranges of thermal activation transport as that of NWs 2 and 3. In **Figure S3** of supporting information, it shows that the segment between electrodes 4-5 in NW5 has maintained a linear $R$–$T$ relation after the AC sweeps on the other segment between electrodes 2-3. These results demonstrate that the AC sweeps can change locally the resistance of a single nanowire, which renders a blue print of selective phase change based nanowire bit-train.

High-resolution TEM analysis on the two segments in NW5 after AC sweep was performed to examine the microstructure, as shown in **Figure 7**. **Figure 7a** shows the SEM image of the to-be-carved-out regions for cross-section TEM analysis, as indicated by yellow and blue arrows. **Figure 7b** shows the cross-section TEM image in the segment between contacts 2-3, which has been subject to AC sweeps. **Figure 7d** and **Figure 7f** are the high resolution TEM and FFT images with hazy ring patterns, indicating that the region D is predominantly amorphous, corresponding to the high resistance and semiconducting behavior. Region B shows the same structural features; whereas, regions A and C show a layered structure, likely to be in transition to amorphous state (refer to **Figure S4** in supporting information.). In contrast, **Figure 7c** shows the cross-section TEM image of the segment between contacts 4-5, which has not been subject to AC sweeps. Apart from the amorphous and layered structures respectively shown in region E and F, region G reveals a polycrystalline structure with clear diffraction spots, as shown in the high-resolution TEM image and FFT image (**Figure 7e** and **Figure 7g**). This region contributes dominantly to the low resistance and metallic conduction behavior. Furthermore, the Sb and Te elemental composition and stoichiometry are measured by EDX in the center of NW5, and the detailed results are shown in the supporting information. Both segments in NW5 have similar EDX spectra (**Figure S5**) and the atomic ratio of Sb:Te is confirmed to be 2:3 (**Figure S6**), which reveals that the nanowire has minimal oxidation or composition change from the AC sweep measurements.

In summary, what we have observed are reproducible sharp increases in resistance at ~3 GHz as a result of AC sweeps on single $Sb_2Te_3$ nanowires. The low to high resistance switch by 6 - 7 orders of magnitude at ~77 K is accompanied by a change of metallic to semiconducting conduction characteristic, and correspondingly from crystalline to amorphous structure. At first glance, one may deduce from these evidences that the MW power at about 1 μW may have induced a phase change within the $Sb_2Te_3$ nanowire, via dielectric heating, comparable to that in the 2.45 GHz microwave oven since the dielectric constant of $Sb_2Te_3$ at 77 K [10] is on the same order of magnitude with that of water at room temperature. Refer to the heat capacity equation from Pashinkin *et al.*,[19] a rough calculation of the energy required for a 2.5 μm long, 200 nm in radius NW segment, to be heated up from 77 to 403 K (130 °C) is ~0.1 nJ. Thus, the 1 μW (-30 dBm) AC voltage could easily provide enough energy to heat up and amorphize the NW. Indeed, such heating has even caused morphology changes shown in NW5 (**Figure 7a**). Nevertheless, this interpretation fails to address the facts that the transition occurs only ~3 GHz, and also the phase change is concentrated locally in a segment within a wire.

Such sharp transition at a resonant frequency signals that the mechanism is of an electronic transition nature. Recent work on bonding calculations and simulations [20-22] suggests that the transition from low resistance to high resistance state in PCMs originate from the change of chemical bonding. Wuttig *et al.* has proposed a new state of matter for materials, such as $Sb_2Te_3$, $Ge_2Sb_2Te_5$, GeTe, *etc.*, so called "incipient metals".[20] Their unique metavalent bonding goes beyond the characteristic parameters for conventional solid state bonding, and possesses bonding mechanism between those of covalency and metallicity, but at the same time, distinctly different from both. We believe that the origin of the phase transition we have observed is induced by the distortion in bonds under charge shuffling resonating at 3 GHz, such as in the resonance-like bonds formed between Sb and Te layers by 5p electrons. As a result, it leads to the transition from delocalized to localized electron distributions,[23,24] rendering self-trap potential wells,[8,25] and exhibiting thermally activated transport as shown in the transport measurements. In depth investigations are required to elucidate the nature of the atomistic bonding mechanisms.

## 4. Conclusion

Reproducible resistance increases in $Sb_2Te_3$ nanowires are observed at ~3 GHz during radio frequency to microwave AC voltage sweeps. The resistance jump is evidenced by a transition from crystalline metallic to amorphous semiconducting phase. This the low to high resistance change is accompanied by a phase change from a rhombohedral crystalline to an amorphous disordered structure, which is verified by high-resolution electron microscopy analysis performed at nanowire segments with and without AC sweeping measurements. Correspondingly, the resistance change manifests in a metallic to semiconducting transition, which is demonstrated in the temperature dependent conductance measurements.

This new phenomenon of microwave AC voltage induced amorphization of crystalline phase change material is of fundamental interest for both experimentalists and theorists to elucidate the physical nature of the phase change, especially from the perspective of bonding mechanisms. In addition, it is also unknown what role the surface states play in the phase change in the $Sb_2Te_3$ nanowire, which is known to have topological surface state conduction. On the other hand, this discovery could bring technology innovation for neuromorphic computing devices. With an additional AC frequency control, one can envision potential multi-state information bit encoding and discrimination along a single nanowire, *i.e.* a phase change nanowire based bit train.

**Supporting Information**
Supporting Information is available from the Wiley Online Library or from the author.

**Acknowledgements**

We also thank Dr. Jinzhong Zhang for technical assistance of e-beam lithography and Dr. Martina Luysberg and Abdur Jalil for high resolution TEM imaging.

JGL is indebted from the insightful discussions with Prof. Matthias Wuttig and Prof. Riccardo Mazzarello.
ST acknowledges supports from the National Science Foundation (DMR-1508661 and CHE-1611134 as well as CHE-2004252 with partial co-funding from the Quantum Information Science program in the Division of Physics).

**Conflict of Interest**

The authors declare no conflict of interest.

**References**

[1] N. Yamada, E. Ohno, K. Nishiuchi, N. Akahira, M. Takao, *J. Appl. Phys.*, http://doi.org/10.1063/1.348620.

[2] A. Kolobov, P. Fons, A. Frenkel, A. Ankudinov, J. Tominaga, T. Uruga, *Nat. Mater.*, http://doi.org/10.1038/nmat1215.

[3] M. Wuttig, N. Yamada, *Nat. Mater.*, http://doi.org/10.1038/nmat2009.

[4] H. -. P. Wong, S. Salahuddin, *Nat. Nanotechnol.*, https://doi.org/10.1038/nnano.2015.29.

[5] F. Rao, K. Ding, Y. Zhou, Y. Zheng, M. Xia, S. Lv, Z. Song, S. Feng, I. Ronneberger, R. Mazzarello, W. Zhang, E. Ma, *Science.*, http://science.sciencemag.org/content/358/6369/1423.abstract.

[6] W. Zhang, R. Mazzarello, M. Wuttig, E. Ma, *Nat. Rev. Mater.*, https://doi.org/10.1038/s41578-018-0076-x.

[7] S. A. Song, W. Zhang, H. Sik Jeong, J. Kim, Y. Kim, *Ultramicroscopy.*, http://doi.org/10.1016/j.ultramic.2008.05.012.

[8] A. Pirovano, A. L. Lacaita, A. Benvenuti, F. Pellizzer, R. Bez, *IEEE Trans. Electron Devices.*, https://doi.org/10.1109/TED.2003.823243.

[9] T. Kato, K. Tanaka, *Jpn. J. Appl. Phys.*, http://doi.org/10.1143/JJAP.44.7340.

[10] O. Madelung, U. Rossler, M. Schulz, *Non-Tetrahedrally Bonded Elements and Binary Compounds*, Springer, Springer-Verlag Berlin Heidelberg **1998**.

[11] Y. C. Arango, L. Huang, C. Chen, J. Avila, M. C. Asensio, Detlev Grützmacher, Hans Lüth, G. L. Jia, Thomas Schäpers, *Sci. Rep.*, https://doi.org/10.1038/srep29493.

[12] M. A. Caldwell, R. G. D. Jeyasingh, H. -. P. Wong, D. J. Milliron, *Nanoscale.*, http://dx.doi.org/10.1039/C2NR30541K.

[13] B. Kersting, M. Salinga, *Faraday Discuss.*, http://dx.doi.org/10.1039/C8FD00119G.

[14] J. S. Lee, S. Brittman, D. Yu, H. Park, *J. Am. Chem. Soc.*, https://doi.org/10.1021/ja711481b.

[15] Da Silva, Juarez L. F., *J. Appl. Phys.*, http://doi.org/10.1063/1.3533422.

[16] Y. Zheng, M. Xia, Y. Cheng, F. Rao, K. Ding, W. Liu, J. Yu, Z. Song, S. Feng, *Nano Res.*, https://doi.org/10.1007/s12274-016-1221-8.

[17] J. Wang, J. Wang, Y. Xu, T. Xin, Z. Song, M. Pohlmann, M. Kaminski, L. Lu, H. Du, C. Jia, R. Mazzarello, M. Wuttig, W. Zhang, *Phys. Status Solidi RRL.*, https://doi.org/10.1002/pssr.201900320.

[18] S. A. Baily, D. Emin, *Phys. Rev. B.*, https://doi.org/10.1103/PhysRevB.73.165211.

[19] A. S. Pashinkin, A. S. Malkova, M. S. Mikhailova, *Russ. J. Phys. Chem. A.*, https://doi.org/10.1134/S0036024408050336.

[20] M. Wuttig, V. L. Deringer, X. Gonze, C. Bichara, J. Raty, *Adv. Mater.*, http://doi.org/10.1002/adma.201803777.

[21] S. Clima, D. Garbin, K. Opsomer, N. S. Avasarala, W. Devulder, I. Shlyakhov, J. Keukelier, G. L. Donadio, T. Witters, S. Kundu, B. Govoreanu, L. Goux, C. Detavernier, V. Afanas'ev, G. S. Kar, G. Pourtois, *Phys. Status Solidi RRL.*, https://doi.org/10.1002/pssr.201900672.

[22] P. Noé, A. Verdy, F. d'Acapito, J. Dory, M. Bernard, G. Navarro, J. Jager, J. Gaudin, J. Raty, *Sci Adv*., https://doi.org/10.1126/sciadv.aay2830.

[23] J. Raty, M. Schumacher, P. Golub, V. L. Deringer, C. Gatti, M. Wuttig, *Adv. Mater.*, http://doi.org/10.1002/adma.201806280.

[24] M. Rütten, A. Geilen, A. Sebastian, D. Krebs, M. Salinga, *Sci. Rep.*, https://doi.org/10.1038/s41598-019-43035-7.

[25] J. L. M. Oosthoek, D. Krebs, M. Salinga, D. J. Gravesteijn, G. A. M. Hurkx, B. J. Kooi, *J. Appl. Phys.*, https://doi.org/10.1063/1.4759239.

Figures

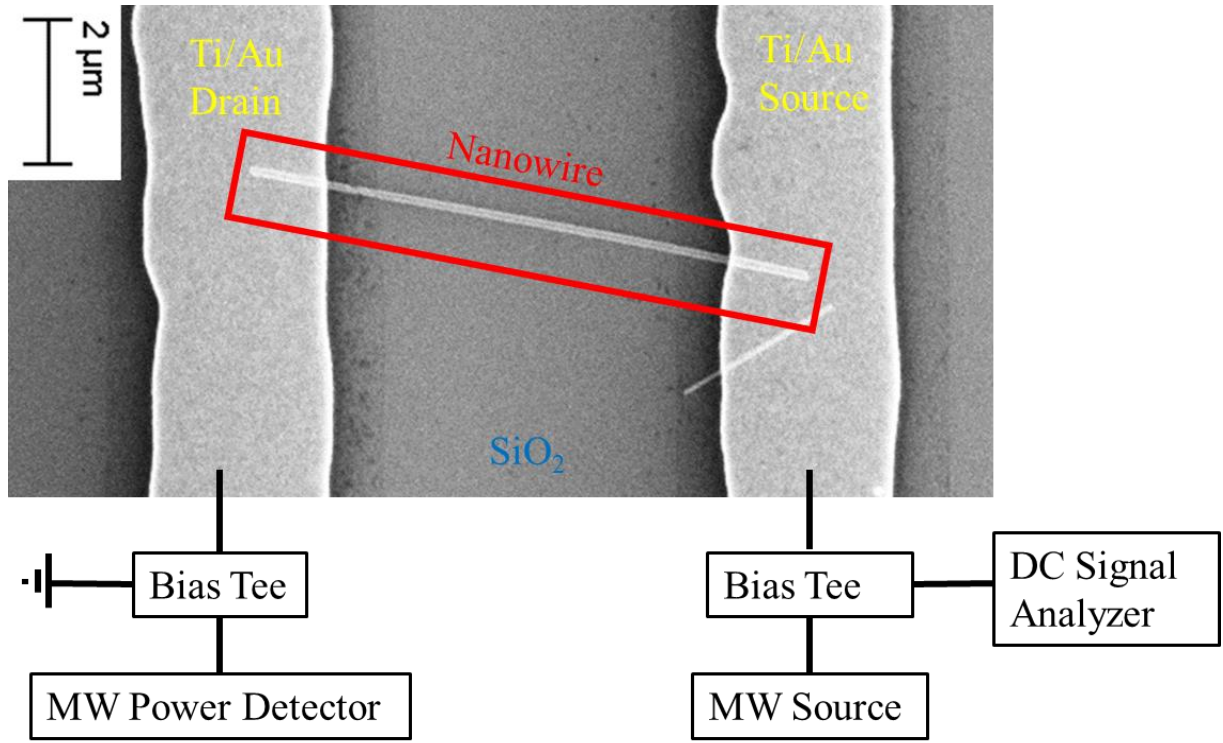

**Figure 1.** SEM image of a contacted single Sb₂Te₃ nanowire along with the measurement circuit diagram.

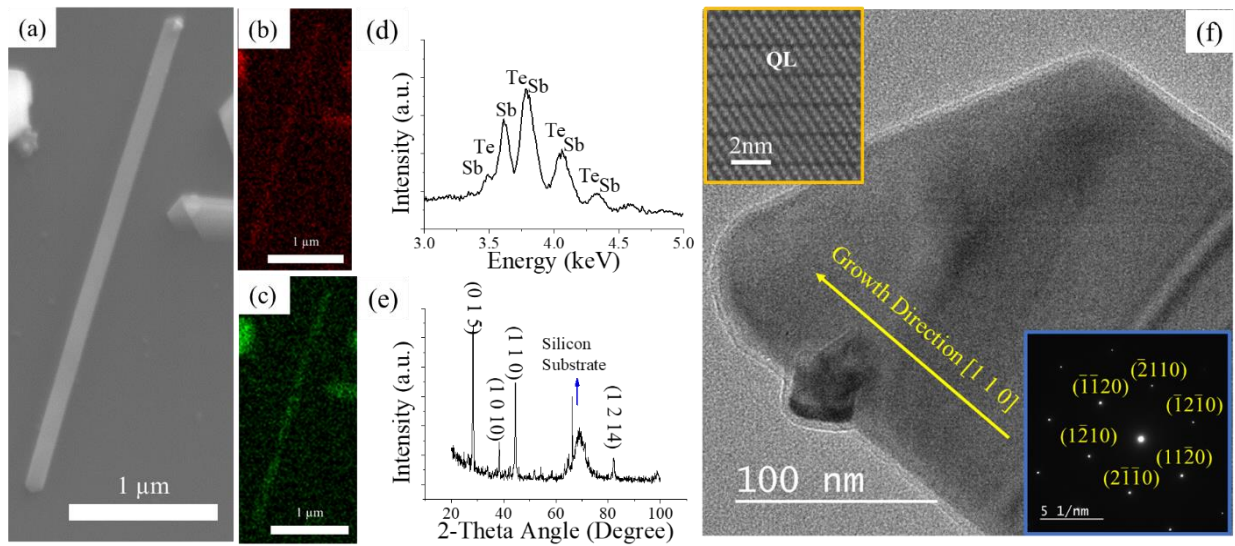

**Figure 2.** (a) SEM image of a Sb₂Te₃ NW on a Si/SiO₂ substrate. Images (b) and (c) illustrate the EDX color mappings of the respective Sb and Te elements. (d) EDX spectra and (e) XRD spectra of Sb₂Te₃ nanowires. The peaks fit with the rhombohedral structure of Sb₂Te₃ (PDF # 00-015-0874). (f) TEM image with lower right SAED inset showing the hexagonal crystalline structure; and upper left cross-section HRTEM inset displaying the stacked quintuple layers (QL).

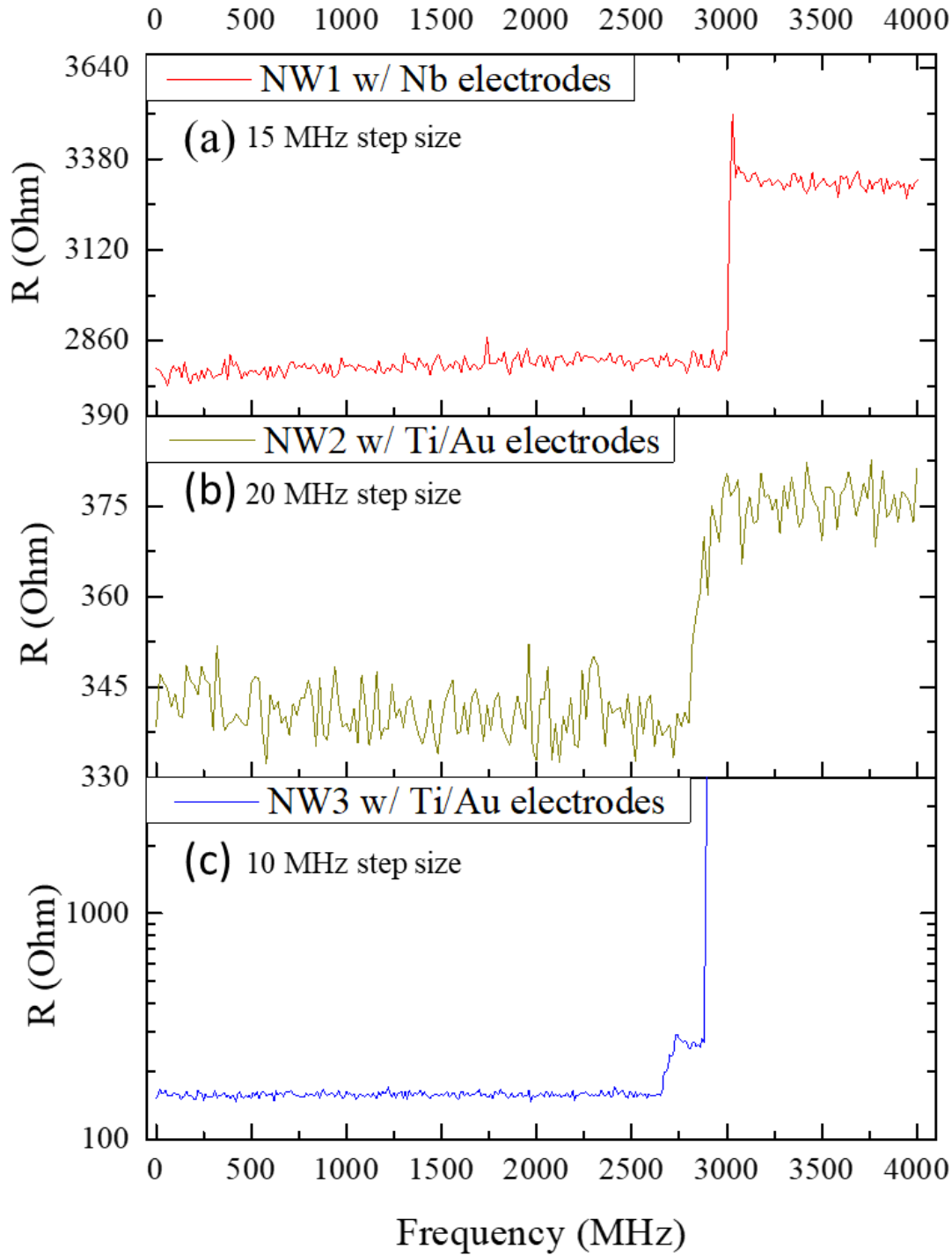

**Figure 3.** Electrical resistance measurements at first forward AC voltage sweeps from 10 MHz to 4 GHz with 0.33 s DC integration time. Consistent increases occur at ~3 GHz for all three NW samples. Forward sweep results for samples NW1, NW2 and NW3 are separately shown in (a), (b) and (c). In (c), the resistance rise at 3 GHz jumps out of the preset range of DC measurement.

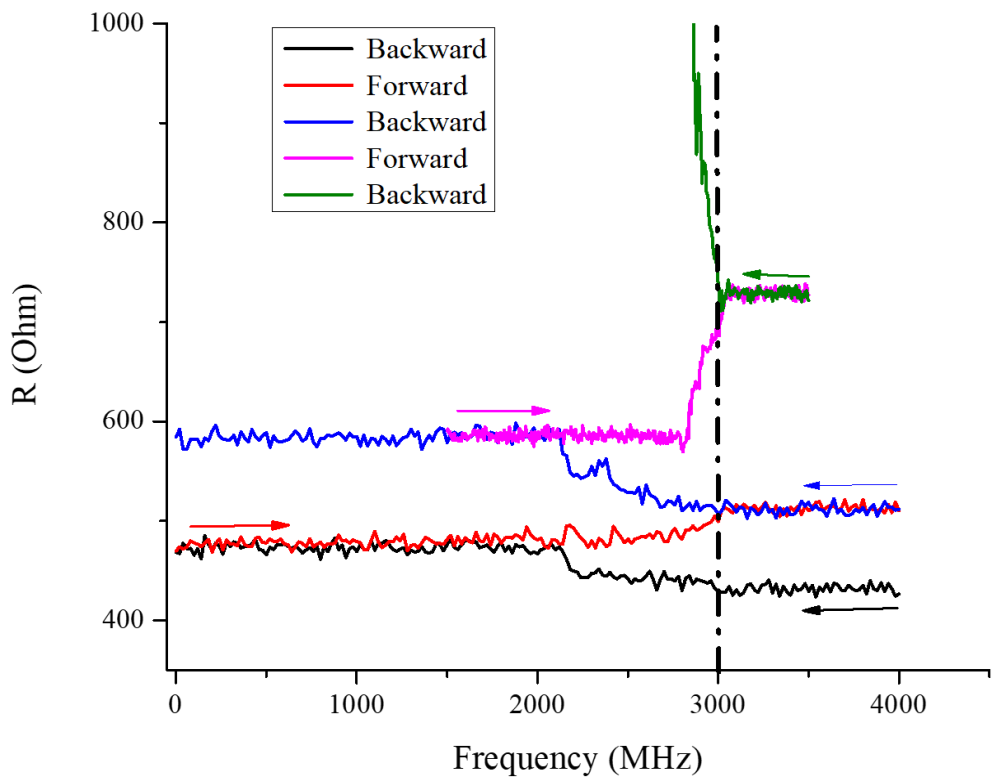

**Figure 4.** Subsequent continuous backward and forward AC voltage sweep results for NW2 at ~1 μW power level. The sweep directional arrows with color correspond to the data plot in the same color.

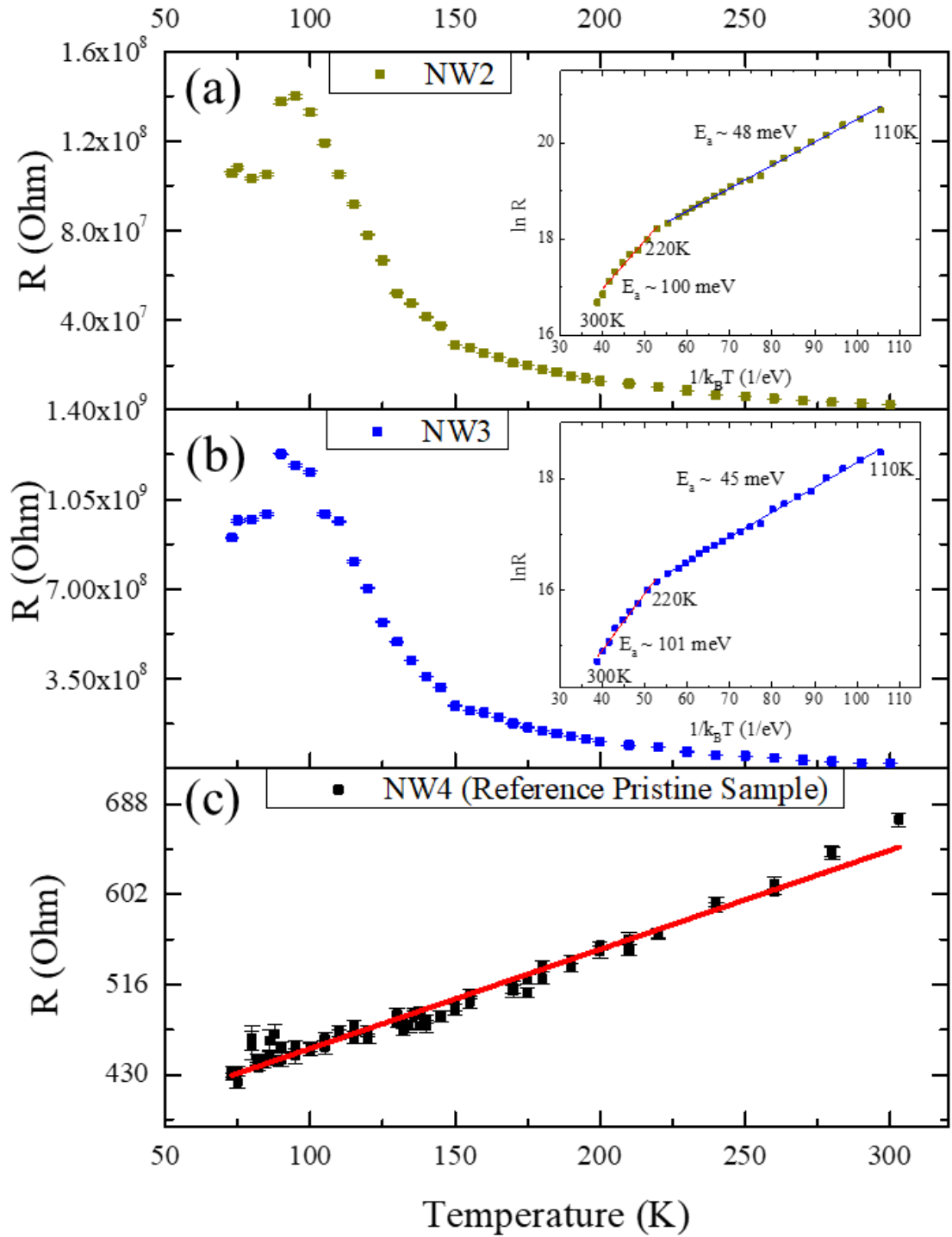

**Figure 5.** Temperature dependent resistance measurements during warm up from 77 K to 300 K of (a) NW2, inset: corresponding Arrhenius semi-log plot with linear fittings, showing two regions of thermal activated transport; (b) NW3, inset: corresponding Arrhenius correlation, suggesting two regions with respective activation energy of ~48 and ~100 meV. (c) NW4 reference sample with linear regression fit in red demonstrating metallic behavior.

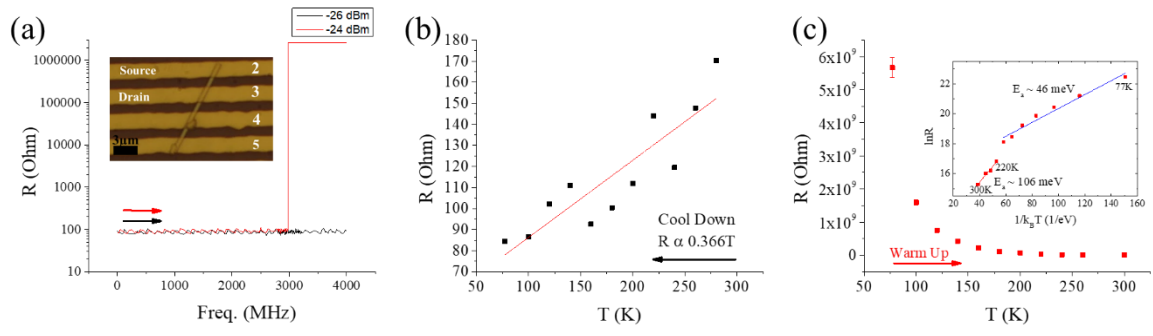

**Figure 6.** (a) AC voltage sweep at 77 K across electrodes 2-3 of NW5. Resistance starts to increase sharply at 3 GHz for -24 dBm sweep (red line). Inset: optical image of NW5 with four electrodes. (b) *R–T* plot at cool down for NW segment 2-3 before AC sweeps, showing linear metallic conducting behavior. (c) *R–T* plot at warm up stage after MW switching to high resistance state, showing exponential drop in resistance with increasing temperature. Inset: Arrhenius plot confirming the two thermal activation energies agreeable with NW2 and NW3.

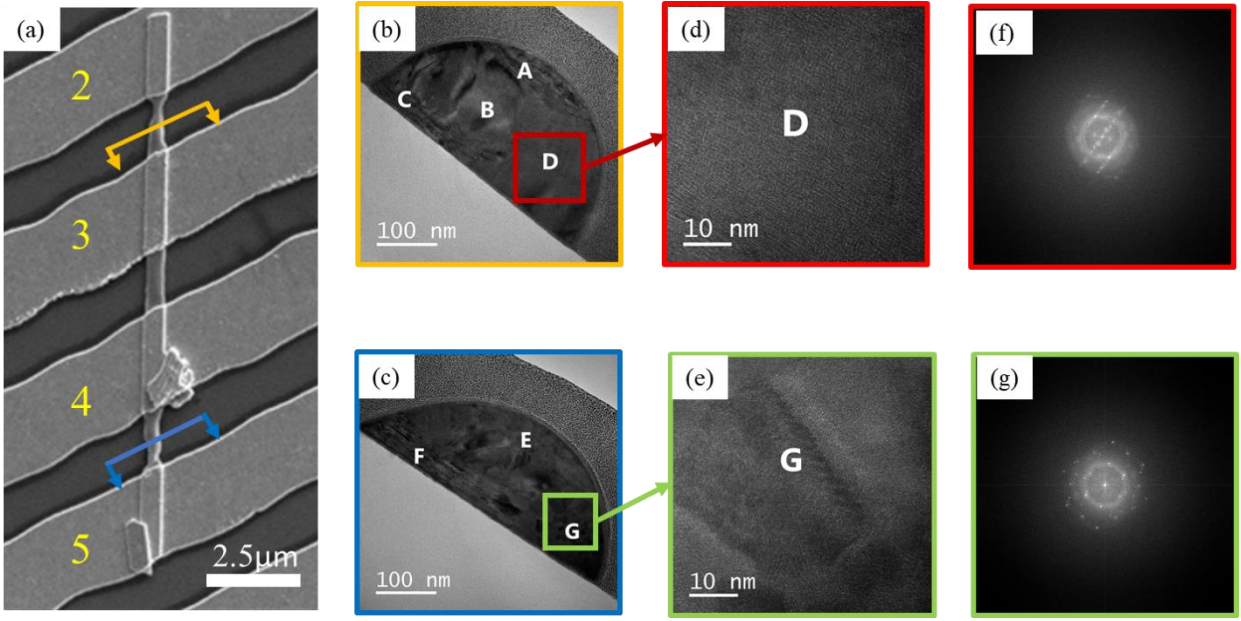

**Figure 7.** TEM cross-sectional analysis of NW5 after AC sweeps. (a) SEM image of NW5 after AC sweep measurements. Yellow and blue arrows indicate the segments selected for TEM cross-section imaging: between contacts 2-3 (underwent AC sweep) and between contacts 4-5 (without AC sweep). (b) TEM cross-section image of a slice of NW5 segment between contacts 2-3, with high resolution TEM and FFT analysis indicating an amorphous phase, as shown in (d & f). (c) TEM cross-section image of a slice of NW5 segment between contacts 4-5, with high resolution TEM and FFT analysis indicating a polycrystalline phase, as shown in (e & g).

P. L. Tse, L. Mugica-Sanchez, F. Tian, O. Rüger, A. Undisz, G. Moethrath, S. Takahashi, C. Ronning and J. G. Lu*

## Microwave AC voltage induced phase change in Sb₂Te₃ nanowires

Cross-examined by charge transport and electron microscopy measurements, the reproducible resistance jump observed at 3 GHz during AC voltage sweeps is evidenced by a transition from crystalline metallic to amorphous semiconducting phase with 6 – 7 orders of magnitude difference in resistance. This is the first demonstration of microwave voltage induced phase change. With this new discovery of AC frequency controlled phase change, one can envision potential multi-bit information encoding along a single nanowire, which could have significant impact for green-IT and neuro-inspired computing devices.

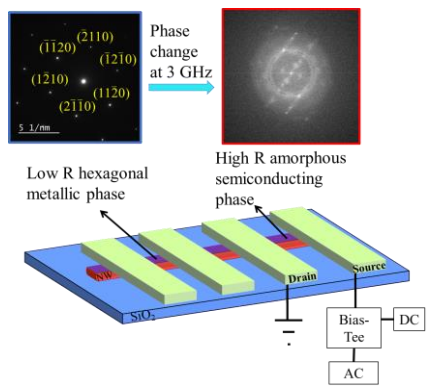

# Supporting Information

**Microwave AC voltage induced phase change in Sb₂Te₃ nanowires**

*P. L. Tse, L. Mugica-Sanchez, F. Tian, O. Rüger, A. Undisz, G. Moethrath, S. Takahashi, C. Ronning and J. G. Lu\**

## S1. Passive heating test on NW2 after AC sweeps

A passive heating test was done on NW2 sample at 130 °C to demonstrate the intrinsic phase transitions mentioned in references [1, 2]. Short 5 s heating and long 60 s heating steps were imposed by placing the NW sample on a piece of aluminum block in the center of a heat plate at 130°C. Upon long 60 s heating on the heat plate at 130 °C, the resistance of NW2, which was driven to a saturation level by the AC voltage sweep, switched back to crystalline phase with resistance in the $10^3$ $\Omega$ range. And with short 5 s passive heating followed by quick quench to room temperature, the nanowire was switched to the amorphous state, resulting in a resistance increase to $10^5$ $\Omega$ range. Passive heating results in **Figure S1** show that the reversible resistance change: decrease to ~4 – 6 k$\Omega$ after applying 60 s long heating, and increase to ~$10^5$ $\Omega$ after 5 s short heating

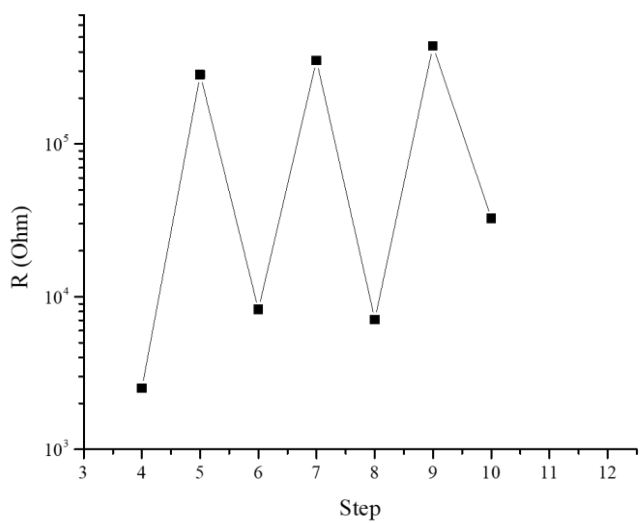

**Figure S1.** Resistance switches under passive heating steps on NW2 at room temperature: 60 s (step 4, 6, 8 and 10) and 5 s (step 5, 7 and 9).

## S2. Saturated high resistance state after phase change

AC sweeps of NW5 segment between contacts 2-3 saturates at high resistance state. **Figure S2** shows the AC sweep result of the segment after it gets switched to high resistance state. The resistance of this segment of NW saturates at ~2 x $10^6$ Ω throughout the range of AC forward and backward sweeps at -24 dBm.

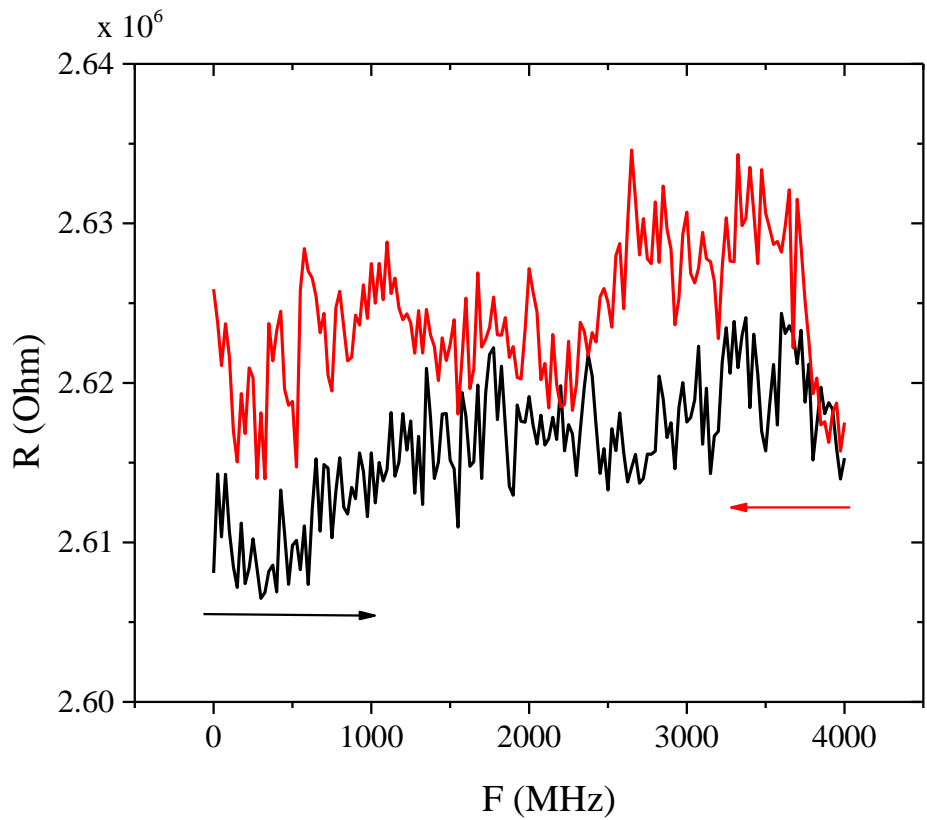

**Figure S2.** AC sweeps of the NW segment between electrodes 2-3 after the NW segment switches and saturates at the high resistance state after continuous sweeps.

## S3. Metallic conduction remain for NW5 segment without AC sweeps

The NW5 segment between contacts 4-5 (not subject to AC sweep) remains in metallic low resistance state. **Figure S3** shows the linear *R–T* plots during cool down and warm up before and after AC sweeps applied to the segment between contacts 2-3.

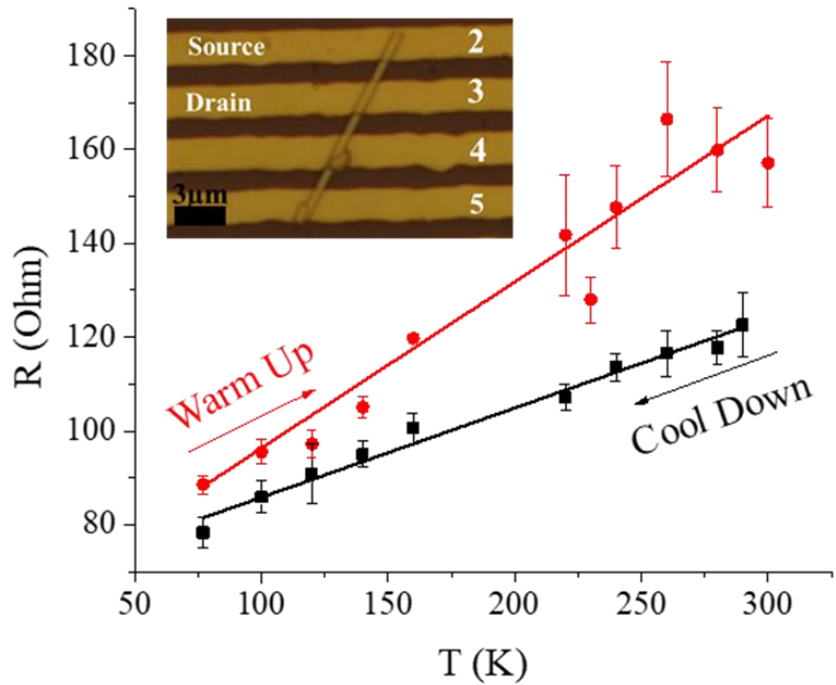

**Figure S3**. Segment of NW5 between electrodes 4-5, before and after AC sweeps carried out in electrodes 2-3, maintains the low resistance metallic state, indicating that information encoding can be done segment-wise along one nanowire. Inset: optical microscope image of the NW5 sample.

**S4. Additional TEM analysis of NW5 segment after undergoing AC sweep**

In addition to the dominate amorphous structure found in this segment between contacts 2-3 of NW5, as explained in the main text, there exists also layered structured region, such as area A indicated in **Figure 7b**, likely to be in transitional phase.

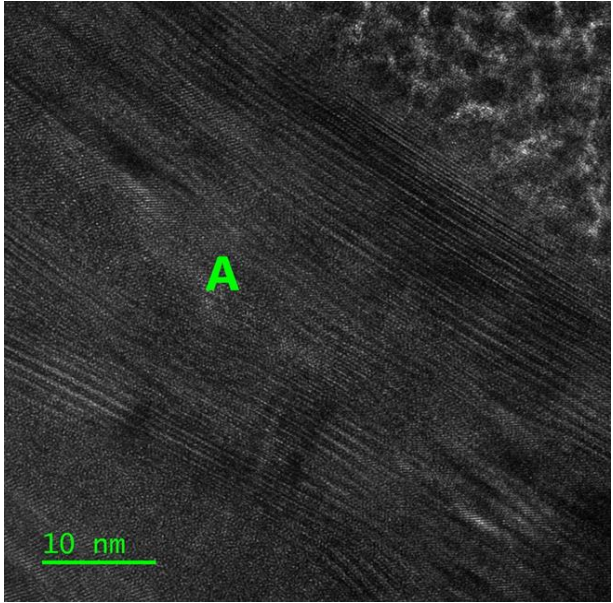

**Figure S4**. TEM cross-section image of area A indicated in **Figure 7b** shows a layered structure.

**S5. Elemental compositions of NW5 segments with and without AC sweeps**

The elemental composition comparison of NW5 was illustrated by carrying out SEM EDX on segments with and without going through AC sweeps. **Figure S5a** shows the SEM image of the cross-section of NW segment between contacts 2-3 which underwent AC sweeps and the EDX measurement at the region marked by the yellow cross is shown in **Figure S5b**. **Figure S5c** shows the SEM image of the cross-section of NW segment between contacts 4-5 without AC sweeps and **Figure S5d** shows the corresponding EDX spectrum. **Figure S5e** shows the overlap of these two EDX spectra, indicating that the two spectra are similar in the signal intensity related to Sb and Te. The results demonstrate that both segments preserve the $Sb_2Te_3$ composition with no oxidation or decomposition.

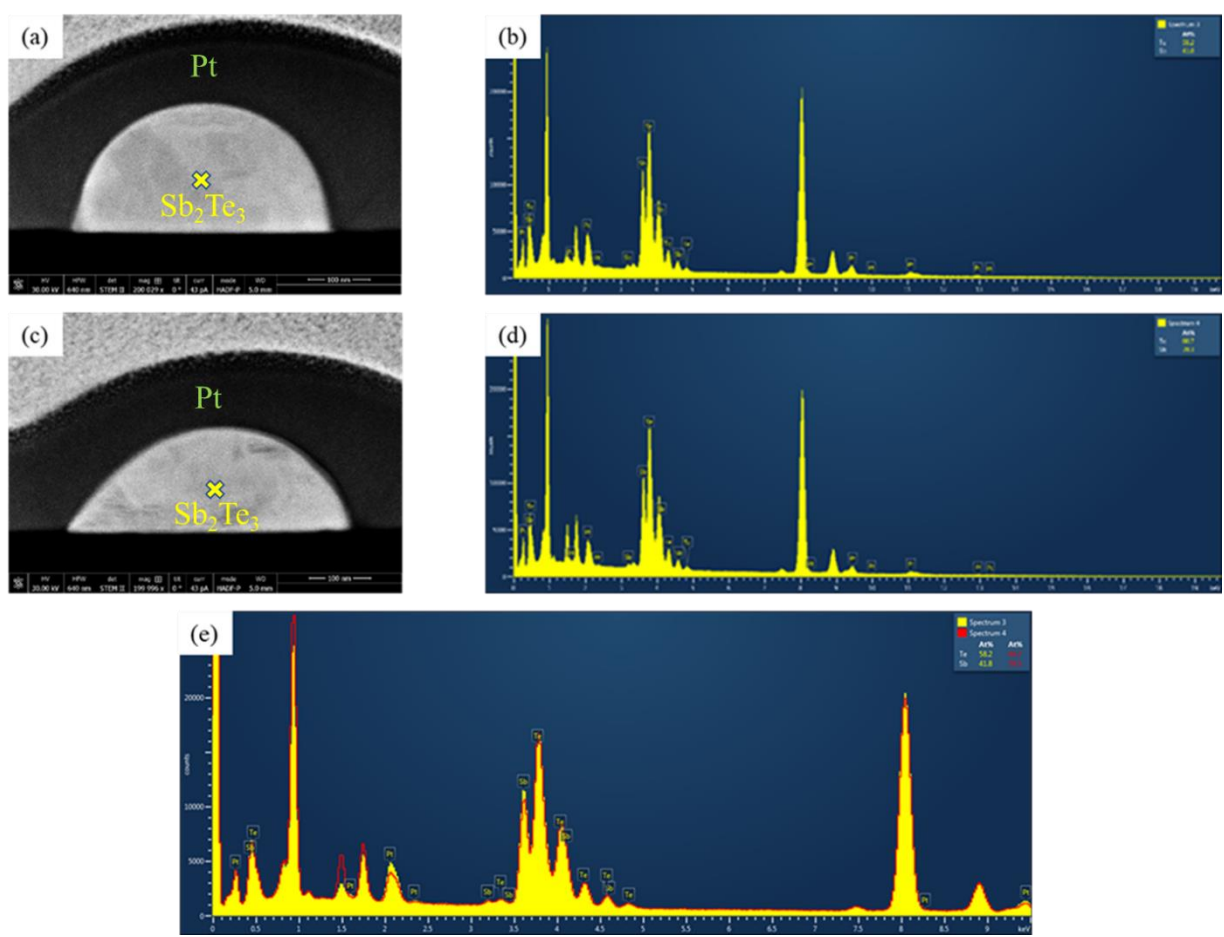

**Figure S5.** Consistent EDX Spectra of NW segments. (a) and (b) SEM image of the cross-section of the NW segment between contacts 2-3 and the corresponding EDX spectrum. (c) and (d) SEM image of the cross-section of the NW segment between contacts 4-5, and its EDX spectrum. (e) Overlap of the two EDX spectra, showing similar Sb and Te contents.

## S6. Elemental composition of NW5 segment by TEM EDX

The atomic ratio of the NW is also measured by the TEM EDX line scan. **Figure S6a** shows the TEM image of the NW segment between contacts 4-5 and the red dash line indicates the region of line scan. **Figure S6b** is the atomic ratio plot along the scan line, showing the atomic percentage of Sb around 40% and Te about 60%, confirming the 2:3 of Sb:Te stoichiometry.

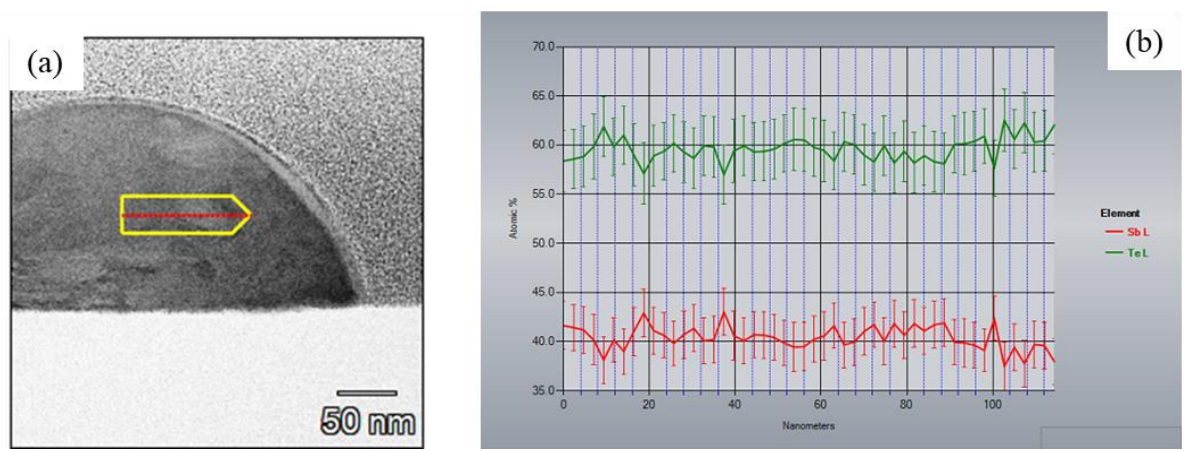

**Figure S6.** EDX lines scan of NW5 at the segment between contacts 4-5. (a) SEM image of the cross-section of the segment, the dash line indicates the line scan region. (b) Atomic ratio along the line scan, with the Sb: Te = 2:3.

## References for supporting information

[1] Y. Zheng, M. Xia, Y. Cheng, F. Rao, K. Ding, W. Liu, J. Yu, Z. Song, S. Feng, *Nano Research.*, https://doi.org/10.1007/s12274-016-1221-8.

[2] H. Choi, S. Jung, T. H. Kim, J. Chae, H. Park, K. Jeong, J. Park, M. Cho, *Nanoscale.*, https://doi.org/10.1039/c6nr05852c.